# Simulating Strategies for Defense Against Brand-Targeted Online Disinformation

Bryan C. Boots
Henry W. Bloch School of Management
University of Missouri-Kansas City
Kansas City, USA
bootsbr@umkc.edu

***Abstract*—*Brand-targeted disinformation can damage firms not only by spreading false beliefs, but by eroding trust, amplifying reputational uncertainty, and persisting across networked digital environments. This paper models brand-targeted disinformation as a trust-weighted contagion process using a multi-agent simulation across six network structures, three amplification levels, and six intervention strategies. Results show that scale-free and influencer-heavy networks are most vulnerable to rapid spread, while clustered and polarized networks provide partial containment until amplification bridges communities. Hybrid intervention combining prebunking, trusted-node activation, and rapid correction performs best overall. The paper contributes a reusable simulation framework for studying active brand disinformation defense strategies.*

## I. Introduction

Online disinformation is no longer only a problem for elections, public health, or geopolitical conflict; it is also an emerging problem for firms, brands, products, and markets [1]. A false claim about a product safety failure, a fabricated customer experience, a misleading allegation about a firm's political activity, or a coordinated campaign of fraudulent reviews can spread through online social networks (OSNs), search systems, recommendation systems, and influencer communities before the targeted firm has a reasonable opportunity to respond. The resulting harm may appear not only as belief in a particular false claim, but also as reputational damage, reduced trust, changes in purchase intention, stakeholder uncertainty, negative algorithmic associations, or the persistence of a damaging narrative long after correction.

The problem is made more difficult because online disinformation does not spread as a simple broadcast from one source to many receivers. It spreads through networks of people, agents, communities, platforms, and trust relationships. Prior work has shown that false information can diffuse differently than true information online, that misinformation networks have identifiable structural and behavioral properties [2], and that coordinated influence campaigns can be mapped as networked phenomena [3]. These findings suggest that brand-targeted disinformation should be understood not simply as a false message, but as a sociotechnical process in which source credibility, network position, timing, reinforcement, and platform visibility jointly shape spread and persistence.

Many existing approaches focus on identifying fake news, misinformation, and disinformation using data mining, natural language processing, machine learning, user features, propagation features, or combinations of these [4, 5]. These methods are necessary, but detection is not the same as containment. Even if a false brand claim is identified, reputational harm depends on who has seen it, who believes it, who shared it, which communities it has entered, which trusted actors have reinforced it, and whether credible actors are available to correct it.

This is especially important because firms occupy an unusual position in the trust network. The targeted firm may possess the most accurate information about the disputed claim, but it is not necessarily the most trusted messenger. In conventional crisis communication, an organization may issue a statement, provide evidence, and attempt to restore confidence [6]. In a networked disinformation setting, however, a corporate rebuttal may be interpreted by skeptical audiences as self-interested or incomplete. This creates a credibility gap that is not easily solved by faster posting or stronger wording.

Brand-targeted disinformation may also create direct economic incentives for manipulation. Prior work on online review fraud shows that reputational manipulation in platformed markets can have measurable competitive and market consequences [7]. Thus, brand disinformation is not only a communication problem; it may also be a market manipulation problem, a platform governance problem, and a Web Intelligence problem. A firm's reputation increasingly exists not only in the minds of customers, but also in the distributed and algorithmically mediated systems through which customers encounter information.

In this paper, we seek to better understand how such disinformation may spread through Web-based social networks, how trust-weighted agent behavior influences that spread, and how different intervention strategies alter diffusion dynamics. To do so, we develop a multi-agent simulation model of brand-targeted disinformation as a trust-weighted contagion process. Agents vary in prior brand attitude, susceptibility, source credibility sensitivity, sharing behavior, and trust in other agents. Networks vary across random, scale-free, small-world, clustered, polarized, and influencer-heavy structures. The model also includes an algorithmic amplification parameter to represent the possibility that highly engaging or controversial content receives additional reach through platform visibility systems.

We use this framework to compare five response strategies, described later in the paper. The purpose is not to predict any single real-world brand attack with high precision, but to identify conditions under which different intervention strategies are likely to be more or less effective.

This paper makes three contributions. It frames brand disinformation defense as a trust-weighted diffusion and intervention problem, rather than only as content classification or post hoc crisis communication. Second, it proposes a reusable multi-agent simulation framework for studying how network structure, source credibility, intervention timing, and platform amplification shape disinformation containment. Third, it connects agent networks, information diffusion, and trust models within a Web Intelligence setting.

The remainder of the paper reviews related work, develops the conceptual model, describes the simulation design, presents experimental conditions and results, and discusses implications for Web Intelligence, brand defense, and future research.

## II. Background and Related Work

Brand-targeted disinformation is related to, but distinct from, the broader problem of online misinformation. In the general misinformation setting, the principal object of harm is often public belief: a person encounters a false or misleading claim, treats it as true, and may then act on it or transmit it. In the brand setting, the harm is broader. A false claim about a firm, product, service, or executive may affect belief, but it may also affect brand trust, purchase intention, investor confidence, employee morale, partner relationships, and perceived legitimacy. The target is not only cognition, but reputation.

Brand disinformation may be strategically useful even when eventually corrected. A false claim may be introduced by a hostile actor, repeated by opportunistic influencers, amplified by skeptical communities, or used by competitors and market actors that benefit from reputational uncertainty. The relevant question is therefore not only whether the original claim is true or false, but how it moves, who carries it, which communities receive it, whether trusted actors reinforce it, and whether it becomes embedded in a broader brand narrative. Prior work has shown that online falsehoods can exhibit distinctive diffusion dynamics [8], and that echo chambers and selective exposure shape how information circulates through social media systems [9]. Brand reputation increasingly exists inside the same Web-based social and algorithmic systems in which misinformation spreads.

Traditional crisis communication offers useful insights, but it is not sufficient for this problem. In brand-targeted disinformation, the firm may not know where the claim originated, which community first adopted it, which nodes sustain it, or which messengers skeptical audiences would believe. The firm may also be disadvantaged because it is the interested party. Even when the brand has accurate information, a corporate rebuttal may be interpreted as self-protective rather than neutral. Thus, the response problem is not only communicative, but also networked and computational.

Many misinformation approaches emphasize detection and classification. These methods are valuable, but they do not fully answer the containment problem. A claim can be detected after it has already reached a large audience. A platform can label a claim without knowing whether that label will be trusted. A firm can respond quickly without reaching the community in which the claim is most strongly reinforced. For this reason, this paper models brand disinformation not only as content, but as a diffusion process involving heterogeneous agents, network structure, trust, and intervention timing.

The language of contagion is useful here, but must be applied carefully. An agent may be susceptible to a false claim, exposed to it, adopt it, share it, be corrected, or later become susceptible again. Yet belief is not the same as exposure. A person may see a claim and reject it, ignore it, believe it without sharing it, or adopt it only after repeated exposure from trusted sources. Complex contagion models therefore provide a stronger foundation than simple epidemic models. In complex contagion, adoption depends not only on contact, but also on reinforcement from multiple or socially meaningful sources [10]. Törnberg extends this logic to fake news, showing how echo chambers and polarized clusters can support misinformation as complex contagion [11]. This is relevant for brand disinformation because reputational claims are often interpreted socially.

Trust is central because it is part of the mechanism by which diffusion occurs. Trust and distrust can propagate through online relationships [12], and computational reputation systems such as EigenTrust show how global reputation may be inferred from local trust relationships [13]. More broadly, trust networks have their own topology, dynamics, and measurement problems [14]. A model of brand disinformation should therefore not treat all exposures as equal. Exposure through a trusted node should have a different effect than exposure through an unknown or distrusted node.

This point is important for corporate response. The brand may be a high-information source, but not a high-trust source. Third-party validators, including expert reviewers, journalists, regulators, community leaders, analysts, employees, customers, or creators, may have greater credibility with specific audiences. In the model developed here, these actors are represented as trusted nodes, allowing us to test whether correction works better when it flows through locally trusted messengers rather than through centralized corporate broadcast.

Correction is also not a simple reversal operation. A false claim may continue to influence belief even after correction, a pattern described as the continued influence of misinformation [15]. For brand disinformation, reactive correction may reduce belief among some agents while leaving residual trust damage or enabling later reinfection. Prebunking and inoculation provide a complementary logic. Inoculation theory suggests that people can become more resistant to persuasion when warned in advance and given weakened examples of misleading arguments [16]. Recent work on misinformation games and prebunking similarly suggests that exposure to manipulation techniques can confer resistance against later misinformation [17]. In the brand setting, however, prebunking also depends on trust: a warning from a distrusted firm may be ignored, while the same warning from credible third parties may increase resistance.

Taken together, this prior work points toward the need for a model that combines network structure, agent heterogeneity, source credibility, correction, and pre-exposure resistance. The simulation developed in this paper builds from these ideas.

## III. Conceptual Model and Research Questions

In this paper, we model brand-targeted disinformation as a trust-weighted contagion process. The model begins from the premise that a false brand claim does not become harmful merely because it exists. It becomes harmful when agents encounter it, assign credibility to it, update their beliefs, and possibly transmit it to others. Diffusion, belief, trust, and sharing are therefore treated as related but distinct processes.

The model separates this process into three linked pathways: exposure, adoption, and sharing. In the exposure pathway, agents encounter brand-related claims through network neighbors and, in some conditions, through algorithmically amplified visibility. This follows network diffusion research showing that the structure of connections among agents shapes who receives information and when [18]. However, exposure is not only a network event. Information diffusion differs across topics and contexts [19], and exposure through multiple social contexts can affect adoption differently than repeated exposure through a single context [20]. The model therefore treats exposure as both structural and socially meaningful.

In the adoption pathway, agents decide whether to believe the false brand claim. Adoption is modeled as a probabilistic function of susceptibility, prior brand attitude, source credibility sensitivity, and social reinforcement. This draws from threshold models of collective behavior, in which actors adopt when perceived social support or exposure exceeds a threshold [21]. In the present model, however, the threshold is shaped not only by the number of exposures, but also by source credibility and the agent's existing orientation toward the brand.

In the sharing pathway, agents decide whether to transmit the claim. Belief and sharing are not identical. People may share inaccurate information for reasons other than belief alone, including failures of attention to accuracy [22].

Trust is a key element of the model, and is treated as a relational property that affects both adoption and correction. Prior work on trust transitivity suggests that trust may move through complex social networks, although it weakens and changes across paths [23]. Research on crowdsourced judgments of source quality also suggests that credibility assessments can help combat misinformation online [24]. Here, trust is modeled as part of the mechanism that determines whether disinformation spreads, stalls, or is corrected.

The model includes several stylized agent classes. Consumer agents represent ordinary users and vary in prior brand trust, susceptibility, credibility sensitivity, and sharing thresholds. Hostile actors initiate or deliberately amplify the false claim. Brand advocates hold favorable prior attitudes toward the brand and may counter-message, although their credibility may be limited outside supportive communities. Institutional validators represent experts, journalists, regulators, reviewers, retailers, community leaders, or trusted creators who may possess credibility with audiences that do not necessarily trust the brand itself. The purpose of these agent types is to capture a central feature of the problem: the same message may have different effects depending on who delivers it.

Finally, the model includes algorithmic amplification as a visibility condition. In some cases, content that generates engagement may be shown to more users, reach more communities, or persist longer in feeds. Evidence from platform studies suggests that algorithmic amplification can alter exposure patterns in consequential ways [25]. In this model, amplification is included as a simplified parameter that increases the reach of high-engagement disinformation, allowing us to examine whether amplification changes the timing and targeting requirements for intervention. This conceptual model leads to the following research questions:

*RQ1: How do network structure and trust-weighted agent behavior affect the spread of brand-targeted disinformation in Web-based social networks?*

*RQ2: Which intervention strategies most effectively reduce belief adoption, sharing, persistence, and reputational harm?*

*RQ3: Under what conditions do prebunking and trusted-node activation outperform reactive corporate correction?*

*RQ4: How does algorithmic amplification alter the timing and targeting requirements for effective brand disinformation containment?*

Together, these questions position brand-targeted disinformation as a problem of agents, networks, trust, and intervention. The purpose of the simulation is to explore how local agent-level rules generate global patterns of reputational harm and containment.

## IV. Simulation Design

To answer the research questions, we develop a discrete-time multi-agent simulation model of brand-targeted disinformation spreading through web-based social networks. The model is not intended to reproduce one real brand controversy, product crisis, or platform event. It allows us to systematically vary agent behavior, trust, network structure, intervention timing, and platform amplification to observe how these factors shape diffusion and containment. This follows the logic of agent-based modeling, in which local rules governing individual agents can generate system-level behavior [26]. As Epstein argues, models are useful not only for prediction, but also for explanation, disciplined thought, and the discovery of possible mechanisms [27].

### *A. Model Architecture*

The simulation is modeled as a graph-based multi-agent system. Each node represents an agent, and each edge represents a possible path through which a brand-related claim may be observed, shared, or corrected. Agents are heterogeneous, and differ in prior brand trust, susceptibility, source credibility sensitivity, reinforcement sensitivity, and likelihood of sharing. This follows the view of intelligent agents as entities that perceive their environment and act according to internal rules or goals [28], and the multi-agent systems tradition in which system-level outcomes emerge from interaction among autonomous or semi-autonomous agents [29].

Each run begins with a network, a population of agents, a false brand claim, and a set of experimental conditions. A small number of hostile actors seed the claim. During each time step, agents may be exposed through neighbors who have shared it, or through algorithmically amplified visibility. Agents then update belief, decide whether to share, respond to active interventions, and update trust in the brand. The process continues until a fixed time horizon is reached or the claim falls below a containment threshold. The simulation is implemented in Python. Network generation, graph storage, and metrics are handled with NetworkX [30]. We do not use a prepackaged diffusion library for the core spread process because the model requires custom trust-weighted transition rules. This allows exposure, belief, sharing, correction, reinfection, and trust updating to be represented as distinct processes. To support replicability, simulation conditions are defined through parameter tables and scenario configurations, random seeds fixed, and conditions are repeated across multiple runs.

### B. Agent States and Parameters

Agents occupy several possible states with respect to the false brand claim. A susceptible agent has not adopted the claim. An exposed agent has encountered it but has not adopted or shared it. A believer (infected agent), has adopted the claim and may share it. Interventions (described later) introduce additional states. A corrected, or recovered, agent no longer believes the claim, although brand trust may not fully return to its prior level. A resistant, or inoculated, agent has reduced susceptibility because of prebunking, prior trust, or warning. A reinfected agent was corrected but later returned to belief after repeated exposure or exposure through a highly trusted source. Agent types and parameters are presented in Table I.

Each agent is assigned parameters that govern exposure and intervention response. Prior brand trust represents the agent's initial attitude toward the brand. Source credibility sensitivity determines how strongly the messenger affects adoption. Susceptibility represents the baseline probability of adopting the claim after exposure. Reinforcement sensitivity captures how repeated exposures increase adoption probability, and sharing threshold represents the belief or activation level required before the agent transmits the claim. The model also includes trust decay and trust recovery parameters. When an agent adopts the false claim, brand trust decreases; when corrected, trust may partially recover, but recovery is not assumed to be complete.

TABLE I. SUMMARY OF DIFFERENT AGENT TYPES AND KEY ASSOCIATED PARAMETERS

| Agent type | Interpretation | Key parameters |
|---|---|---|
| consumer | Ordinary user/customer | Medium brand trust; variable susceptibility; ordinary sharing threshold |
| skeptic | Predisposed critic or low-trust consumer | Lower brand trust; higher susceptibility to negative claims |
| brand_advocate | Brand-favorable user | Higher brand trust; lower susceptibility; may counter-message |
| hostile_actor | Attacker or deliberate amplifier | Seeds the false claim; high sharing probability; low brand trust |
| influencer | High-reach account | High degree or visibility; elevated sharing probability |
| institutional_validator | Trusted third party | High correction credibility; used in trusted-node activation |

### C. Network Structures

We simulate diffusion across six synthetic network structures. Synthetic networks allow the effect of topology to be isolated while other conditions are held constant. The first condition is a random network, which serves as a baseline. The second is a scale-free network, representing systems in which a small number of highly connected nodes have disproportionate reach, following the preferential attachment logic described by Barabási and Albert [31]. The third is a small-world network, which combines local clustering with short path lengths and is useful for representing social environments where information can circulate locally while still moving quickly across the broader network [32]. The fourth is a clustered community network, representing affinity groups, fandoms, customer communities, or issue publics. The fifth is a polarized or modular network, in which communities have dense internal ties and fewer cross-community connections. Stochastic blockmodels provide one basis for generating these clustered and modular structures [33]. The sixth is an influencer-heavy network, designed to represent settings where a small number of creators or public accounts dominate visibility. The different simulated network structures are presented visually in Fig. 1.

### D. Algorithmic Amplification

The model includes an algorithmic amplification parameter to represent the possibility that engaging or controversial content receives additional visibility. In the low-amplification condition, agents are exposed primarily through network neighbors. In the moderate-amplification condition, highly shared claims receive additional reach beyond immediate neighbors. In the high-amplification condition, the false claim receives substantially increased reach once sharing or engagement crosses a specified threshold. This parameter does not attempt to reproduce a full ranking or recommendation system. Under amplification, content can jump beyond local ties and reach agents who otherwise might not encounter it. In the brand disinformation context, this means that algorithmic visibility may convert a local reputational attack into a broader market-facing event.

## V. EXPERIMENTAL CONDITIONS AND OUTCOME METRICS

The simulation experiments compare how intervention strategies perform under different network, trust, and amplification conditions. The purpose is to observe how intervention effectiveness changes when the same false brand claim spreads through different network structures, visibility environments, and trust distributions. This approach is consistent with work on misinformation containment and influence blocking, where timing and placement of counter-messaging can substantially affect diffusion outcomes [34, 35].

### A. Intervention Strategies

We compare six intervention conditions. The first is a no-intervention baseline, in which hostile actors seed the false brand claim and the simulation proceeds without prebunking, correction, trusted-node activation, or corporate rebuttal. This condition estimates uncontrolled spread and associated brand-trust decline. These strategies are summarized in Table II.

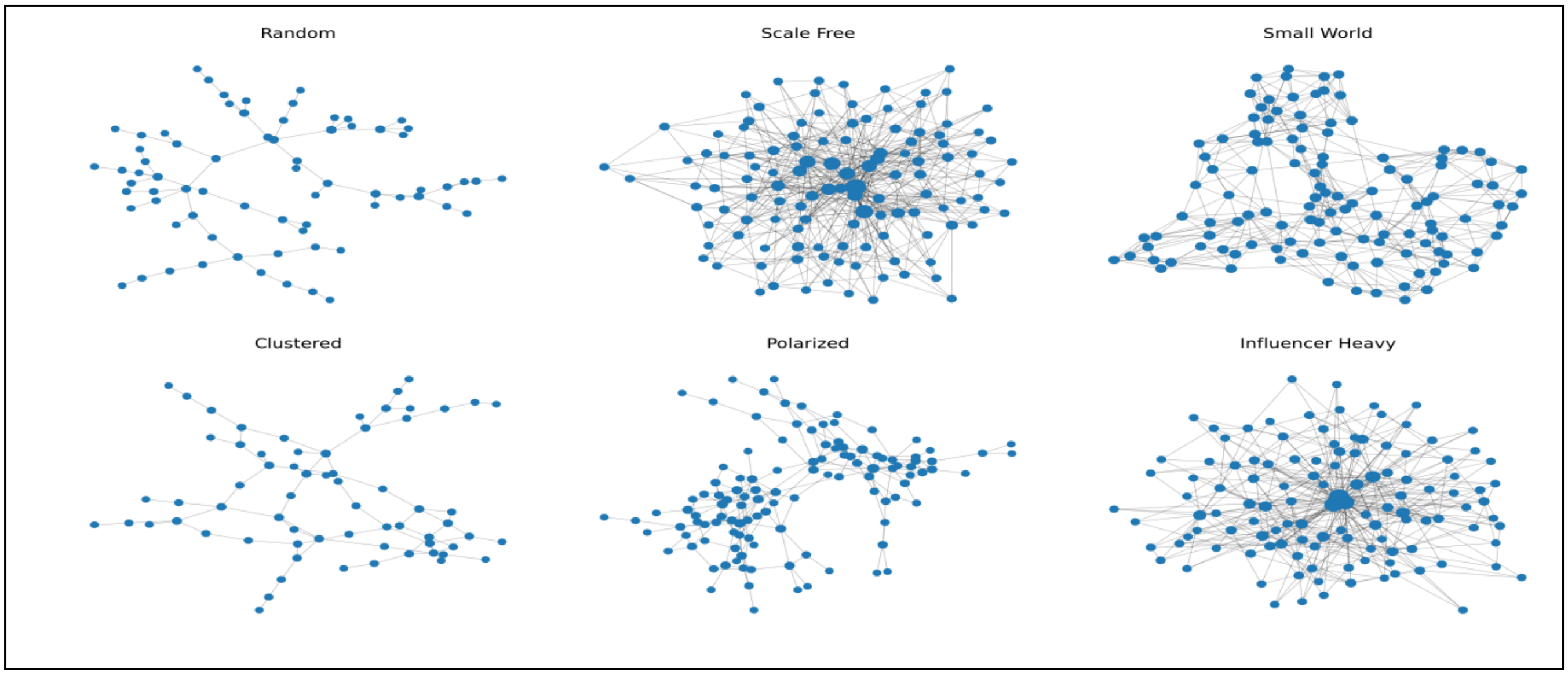

Fig. 1. Network structures generated for use in the simulations

The second condition is reactive correction. Here, the brand issues a corrective message only after the false claim reaches a specified diffusion threshold, such as a proportion of agents exposed, a proportion believing the claim, or a minimum number of sharing events. Reactive correction is modeled as a corporate-originated message with high informational accuracy but variable credibility. This reflects the possibility that audiences may discount corrections from the interested party. Because false claims may continue to influence belief after retraction, correction is modeled probabilistically rather than as guaranteed reversal [15, 36].

The third condition is prebunking. Some agents are made more resistant before the false claim is introduced, representing prior warning, consumer education, transparent communication, or exposure to weakened examples of manipulation techniques. Prebunked agents receive lower susceptibility values and, in some conditions, higher responsiveness to later correction. This follows inoculation and prebunking research suggesting that prior exposure to manipulation techniques can reduce later susceptibility [17]. It also resembles network immunization, where the purpose is to reduce spread before an attack reaches full scale [37].

The fourth condition is trusted-node activation. Correction or counter-messaging is introduced through high-trust third-party validators rather than through the brand alone. Their messages have stronger correction effects among agents who trust them than would be expected from a general corporate rebuttal; this follows research on expert correction [38] and influence blocking, where counter-influence is strategically placed to limit harmful diffusion [35].

The fifth condition is broadcast corporate rebuttal. Unlike reactive correction, which begins after a diffusion threshold is reached, broadcast rebuttal is modeled as a broad brand response once the firm detects the attack. It has greater reach than targeted trusted-node activation, but lower credibility among skeptical agents.

The sixth condition is a hybrid strategy. Prebunking occurs before the attack, trusted nodes are activated as the claim begins to spread, and rapid correction is deployed before the late-stage threshold used in the reactive correction condition. This represents the most prepared organizational condition and tests whether combining resistance, credible messengers, and speed produces better outcomes than any one intervention alone.

### *B. Targeting Strategies*

For interventions requiring agent selection, we compare several targeting strategies. Degree-based targeting selects highly connected nodes, following influence maximization logic that some nodes disproportionately affect diffusion because of their network position [18, 39]. Bridge-node targeting selects agents who connect otherwise separate communities. Community-based targeting selects validators within each cluster, reflecting the idea that trust may be local rather than global. Untargeted broadcast sends correction broadly without regard to network position, providing a realistic but potentially inefficient comparison condition.

TABLE II. SUMMARY OF DIFFERENT STRATEGY INTERVENTION TYPES

| Intervention | Short name | Operationalization |
|---|---|---|
| none | No intervention | Estimates uncontrolled spread and reputational damage |
| reactive_correction | Corporate correction after threshold | Correction activates only after diffusion reaches a threshold |
| prebunking | Pre-exposure resistance | Selected agents begin as resistant or less susceptible |
| trusted_node_activation | Third-party validators | Trusted nodes send correction to their neighborhoods/communities |
| broadcast_corporate_rebuttal | Broad corporate rebuttal | High reach but lower credibility among skeptical agents |
| hybrid | Prebunking, trusted nodes and rapid correction | Preparedness strategy combining resistance, credibility, and speed |

## C. Outcome Metrics

We evaluate outcomes in three categories. Diffusion outcomes include peak infection rate, total exposure, total sharing, and time to peak. Peak infection rate is the maximum proportion of agents who believe the false claim at any point. Total exposure is the proportion who encounter the claim, whether or not they believe it. Total sharing is the proportion who transmit it. Time to peak records how many simulation steps are required for belief to reach its maximum level. These simulation performance metrics are summarized in Table III.

Containment outcomes include time to containment, recovery rate, and reinfection rate. Time to containment measures how long it takes for belief or sharing to fall below a defined threshold. Recovery rate is the proportion of agents who previously believed the claim but are later corrected. Reinfection rate is the proportion of corrected agents who return to belief after further exposure, which is important because correction does not necessarily end the life of a false brand narrative.

Brand-reputation outcomes include trust loss, trust recovery, and cross-community spread. Trust loss captures the aggregate decline in brand trust from the beginning to the lowest point of the simulation. Trust recovery measures the degree to which aggregate trust returns after intervention. Cross-community spread measures whether the claim escapes its original community and enters other clusters. These metrics evaluate not only whether the claim spreads, but whether it persists, whether it is contained, and whether the brand recovers.

# VI. Simulation Results

In this section, we present the simulation results. The reported experiments use 1,000 agents, 20 time steps, and five runs per condition. The full experiment set includes six network structures, six intervention strategies, and three amplification levels, producing 540 simulation runs. Because the number of runs is modest, the results should be interpreted as directional rather than conclusive. Nonetheless, the patterns provide useful early evidence about how network structure, amplification, and intervention strategy interact in brand-targeted disinformation. The results are summarized in Table IV and Figure 2.

TABLE III. Explanation of different metrics used in simulations

| Metric | Definition |
|---|---|
| peak_infection_rate | Maximum proportion of agents believing the false claim |
| total_exposed | Proportion of agents that ever encountered the claim |
| total_believers | Proportion of agents that ever believed the claim |
| total_sharers | Proportion of agents that ever shared the claim |
| time_to_peak | Time step when belief reaches its maximum |
| time_to_containment | First time step after peak when belief falls below threshold |
| recovery_rate | Share of ever-believers who become corrected |
| reinfection_rate | Share of corrected agents who later return to belief |
| trust_loss | Aggregate decline in mean brand trust from initial level to minimum |
| trust_recovery | Increase in mean brand trust from minimum to final level |
| cross_community_spread | Fraction of detected communities reached by the claim |

TABLE IV. Summary of performance of different strategy interventions across key metrics

| Intervention | peak_infection_rate | total_exposed | total_believers | total_sharers |
|---|---|---|---|---|
| broadcast_corporate_rebuttal | 0.465 | 0.866 | 0.679 | 0.573 |
| hybrid | 0.242 | 0.714 | 0.361 | 0.180 |
| none | 0.823 | 0.945 | 0.823 | 0.775 |
| prebunking | 0.789 | 0.932 | 0.789 | 0.743 |
| reactive_correction | 0.345 | 0.848 | 0.600 | 0.475 |
| trusted_node_activation | 0.492 | 0.827 | 0.558 | 0.372 |

Across all network structures and amplification levels, the no-intervention baseline produced a mean peak infection rate of 0.823, with mean total exposure of 0.945 and mean total sharing of 0.775. Thus, without intervention, the false brand claim reached almost the entire simulated network and was shared by more than three quarters of agents on average. Prebunking alone produced only a modest reduction, with a mean peak infection rate of 0.789, total exposure of 0.932, and total sharing of 0.743.

The strongest overall performance was observed in the hybrid intervention condition. Across all settings, hybrid produced a mean peak infection rate of 0.242, mean total exposure of 0.714, and mean total sharing of 0.181. It also produced the highest recovery rate, 0.999, the lowest reinfection rate among active correction strategies, 0.061, and reliable containment in 89 of 90 runs. Reactive correction produced the second-lowest mean peak infection rate, 0.345, but containment was observed in only one run and the mean reinfection rate was 0.377. Broadcast corporate rebuttal produced a mean peak infection rate of 0.465, while trusted-node activation produced a slightly higher value of 0.492 but lower total sharing, 0.372, and much lower reinfection, 0.068. These results show that intervention performance depends on the outcome measure. Reactive correction performed well on peak infection, while trusted-node activation performed better on recovery, reinfection, and containment. Hybrid performed best across the broadest set of measures.

## A. Results by Network Structure

Network structure substantially affected diffusion. Across all interventions and amplification levels, the highest mean peak infection rates occurred in scale-free and influencer-heavy networks, with values of 0.652 and 0.653. Random networks followed at 0.557, small-world networks at 0.511, clustered networks at 0.425, and polarized networks at 0.358. The no-intervention condition showed these differences more clearly. Peak infection reached 0.965 in scale-free networks, 0.947 in influencer-heavy networks, and 0.896 in small-world networks. Lower values were observed for random networks (0.814), clustered networks (0.661), and polarized networks (0.657).

This pattern is consistent with expectations from network science literature. Scale-free and influencer-heavy networks contain highly connected nodes that can accelerate diffusion [31]. Small-world networks combine local clustering with short path lengths, supporting rapid spread across the network [32].

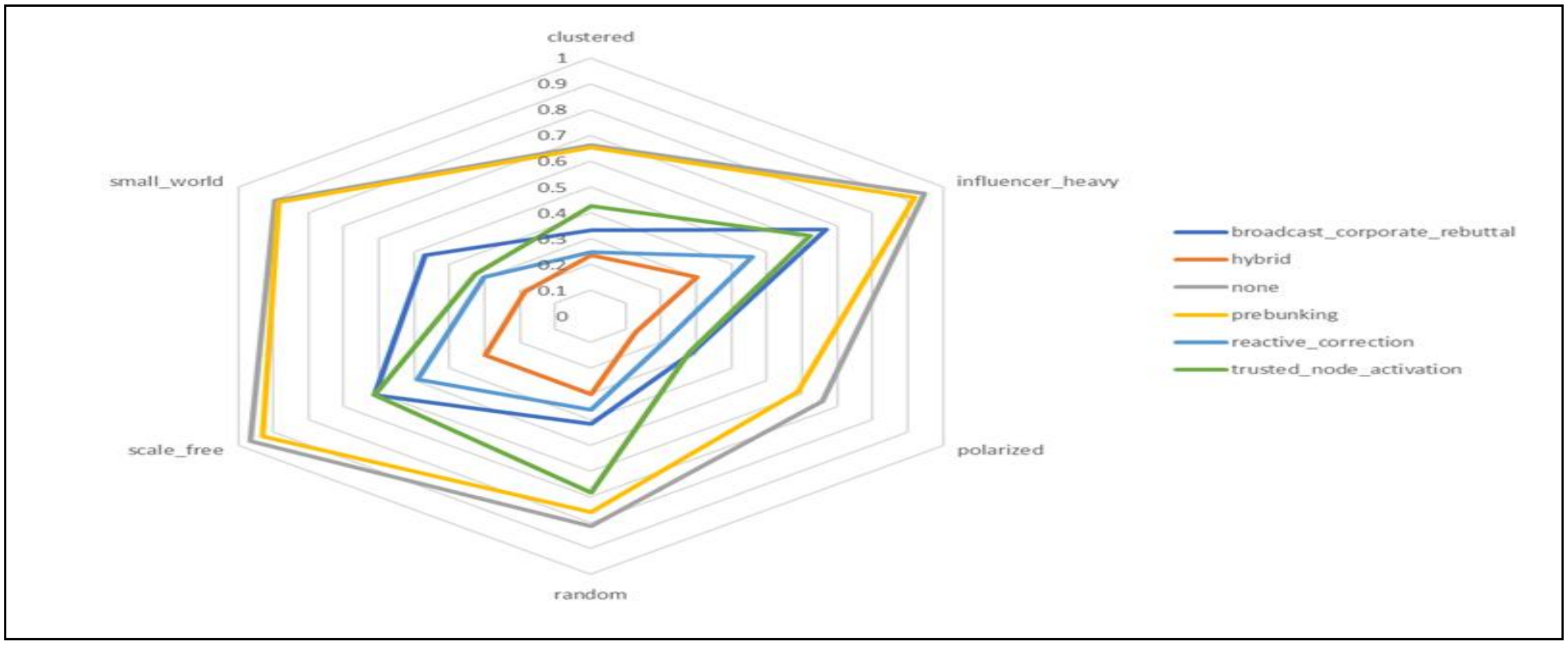

Fig. 2. Results comparison, mean peak infection rate across different network types and intervention strategies

Clustered and polarized networks contain stronger community boundaries, which can slow global diffusion [40, 41].

Hybrid intervention reduced peak infection in all six network conditions. In small-world networks, peak infection fell from 0.896 under no intervention to 0.187 under hybrid intervention. In scale-free networks, it fell from 0.965 to 0.300, and in influencer-heavy networks from 0.947 to 0.301. Hybrid also reduced peak infection to 0.235 in clustered networks and 0.126 in polarized networks.

Reactive correction was generally the next strongest peak-reduction strategy, particularly in scale-free, influencer-heavy, and small-world networks.

Trusted-node activation showed more mixed results: it was relatively effective in small-world and polarized networks, but performed poorly on peak infection in random networks, where it produced a mean value of 0.683. However, in random networks it still produced full recovery and no reinfection, suggesting that it functioned more as a post-peak correction mechanism than as a peak-reduction mechanism.

### B. *Amplification Effects*

Amplification increased spread across the simulation set (see Fig. 4). Across all network structures and interventions, mean peak infection rose from 0.492 under low amplification to 0.521 under moderate amplification and 0.565 under high amplification. Mean total exposure increased from 0.784 to 0.853 and then to 0.930. Total sharing followed the same pattern.

The amplification effect was strongest when intervention was weak or absent. In the no-intervention condition, peak infection increased from 0.764 under low amplification to 0.895 under high amplification. In the prebunking condition, it increased from 0.720 to 0.864. By contrast, hybrid was least affected by amplification, increasing only from 0.231 under low amplification to 0.254 under high amplification.

The effect of amplification also varied by network structure. In the no-intervention condition, clustered networks increased from 0.546 under low amplification to 0.827 under high amplification, and polarized networks increased from 0.548 to 0.778. Random networks increased from 0.759 to 0.898. Scale-free networks showed a smaller increase, from 0.960 to 0.978, and influencer-heavy networks increased from 0.926 to 0.952, largely because these networks were already near saturation under low amplification.

These findings are consistent with the view that amplification can alter effective exposure pathways in online systems [25, 42]. In these simulations, amplification increased peak infection, exposure, and sharing, but the magnitude of the effect depended on the underlying network structure. Amplification had the largest effect in networks where community boundaries otherwise limited global spread, and the smallest effect in networks already dominated by highly connected nodes.

TABLE V. MEAN PEAK INFECTION RATE BY NETWORK TYPE AND INTERVENTION STRATEGY

| | Intervention | | | | | |
|---|---|---|---|---|---|---|
| **Network Structure** | **broadcast** | **hybrid** | **none** | **prebunking** | **reactive** | **trusted_node** |
| clustered | 0.331 | 0.235 | 0.661 | 0.653 | 0.246 | 0.426 |
| influencer_heavy | 0.669 | 0.301 | 0.947 | 0.919 | 0.460 | 0.623 |
| polarized | 0.288 | 0.126 | 0.657 | 0.588 | 0.208 | 0.280 |
| random | 0.418 | 0.303 | 0.814 | 0.759 | 0.364 | 0.683 |
| scale_free | 0.614 | 0.300 | 0.965 | 0.931 | 0.492 | 0.612 |
| small_world | 0.471 | 0.187 | 0.896 | 0.884 | 0.301 | 0.327 |

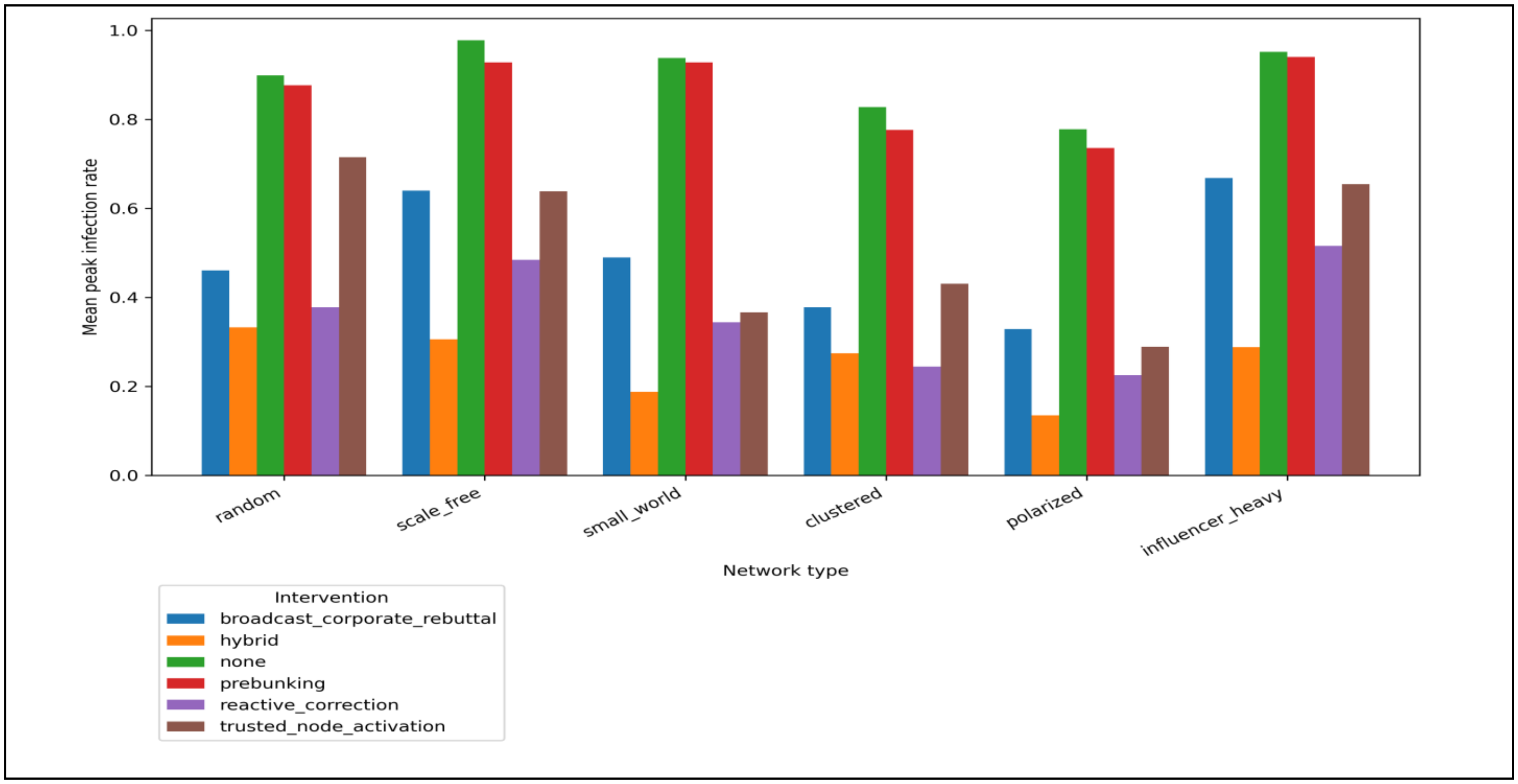

Fig. 3. Peak Infection by Network Type and Intervention Strategy

### C. *Summary of Results*

Taken together, the results show four main patterns. First, network structure shaped the spread of the false brand claim: scale-free and influencer-heavy networks were most vulnerable to near-saturating spread, while clustered and polarized networks produced lower global peak infection. Second, intervention strategy mattered, but different strategies performed best on different outcomes. Third, hybrid intervention consistently produced the strongest overall results. Fourth, amplification increased spread most strongly in networks where structural boundaries otherwise provided some containment. These results provide the empirical basis for the discussion that follows.

## VII. Discussion

The preliminary results show that brand-targeted disinformation defense is better understood as a trust-weighted diffusion and intervention problem in which response effectiveness depends on timing, network structure, messenger credibility, and platform amplification. Detection may tell a firm, platform, or analyst that a claim is false, but it does not show whether the claim has reached vulnerable communities, whether trusted agents have reinforced it, whether correction will be believed, or whether the narrative will continue to circulate after an official rebuttal.

The strongest evidence for this framing is the performance of the hybrid strategy. Across network types and amplification levels, hybrid intervention produced the lowest peak infection, lowest sharing, highest recovery, and most reliable containment. This finding is useful because each component of the hybrid strategy solves a different part of the problem. Prebunking reduces susceptibility before exposure, trusted-node activation improves credibility and recovery, and rapid correction reduces the time available for the false claim to spread unchecked. Together, these mechanisms operate across multiple stages of the diffusion process.

RQ1 asked how network structure and trust-weighted agent behavior affect the spread of brand-targeted disinformation. The simulations show that network structure changes both the scale and speed of diffusion. Scale-free and influencer-heavy networks were most vulnerable to near-saturation, while clustered and polarized networks produced lower global peak infection. This highlights a specific brand-defense problem: when a false claim enters through, or quickly reaches, highly connected agents, the brand has little time to respond before the claim becomes widely visible. At the same time, trust changes the meaning of exposure. Exposure from a trusted agent is not equivalent to exposure from an unknown or distrusted source. Trust is therefore not only an outcome of disinformation; it is part of the mechanism by which disinformation spreads.

RQ2 asked which intervention strategies most effectively reduce belief adoption, sharing, persistence, and reputational harm. The answer depends on the metric. Reactive correction performed relatively well on peak infection, but trusted-node activation performed better on reinfection, recovery, and containment. This suggests that brands should not evaluate response effectiveness only by whether an official correction reduces immediate belief. They should also consider whether the response stabilizes the network after the initial peak. This connects directly to prior work showing that trust and distrust can propagate through networks [12], that trust networks have measurable topology and dynamics [14], and that broker nodes can play an important role in misinformation trust networks [43]. Trusted nodes are part of the credibility infrastructure through which correction becomes believable, and this is consistent with work on source-quality judgments and expert correction [24, 38].

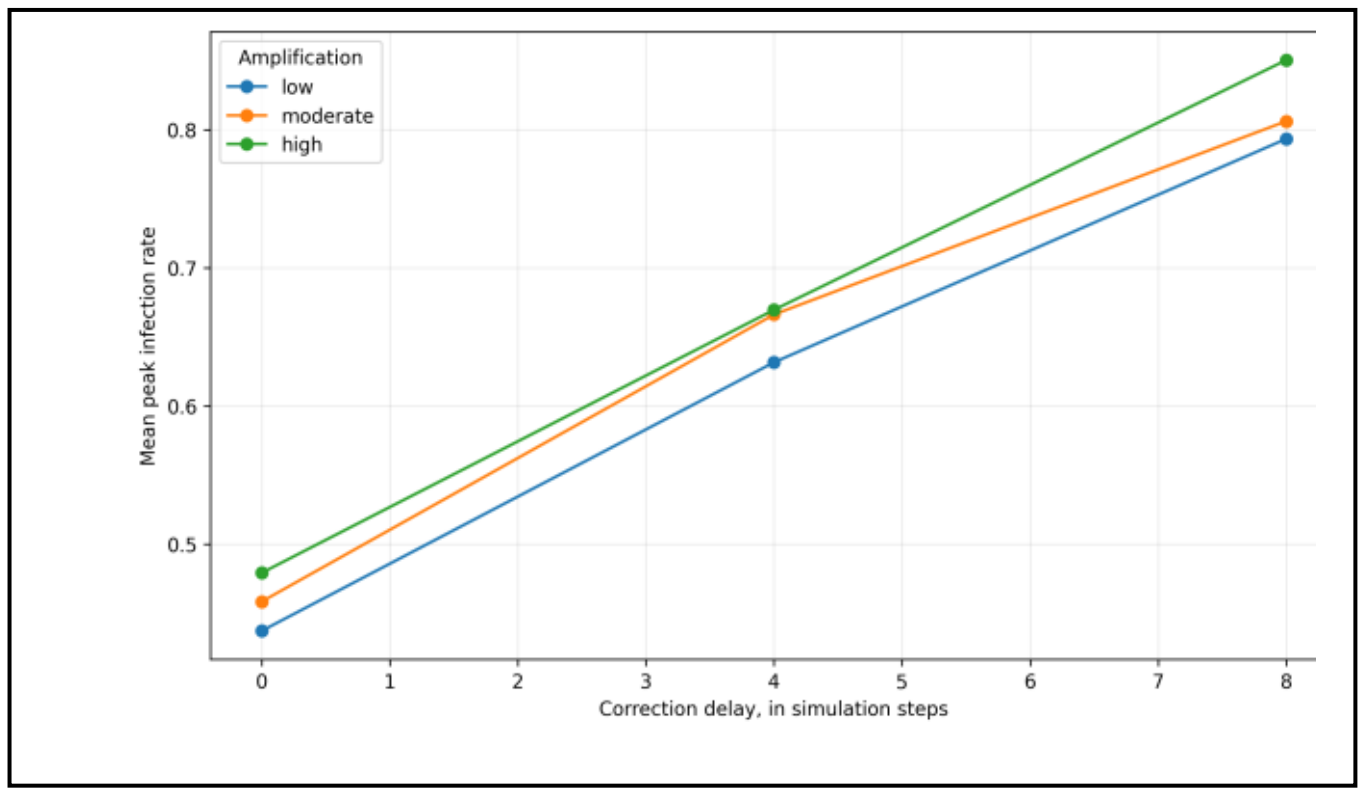

Fig. 4. Amplification sensitivity

RQ3 asked when prebunking and trusted-node activation outperform reactive corporate correction. The answer is mixed. Prebunking alone did not strongly outperform other strategies, especially under high amplification and in hub-dominated networks. This suggests that prebunking may require sufficient coverage, credible delivery, and reinforcement through later interventions. In the hybrid condition, prebunking appears more valuable because it operates alongside trusted-node activation and rapid correction. Trusted-node activation outperformed reactive correction most clearly on recovery, reinfection, and containment, not necessarily on peak infection. Thus, reactive correction may be better for reducing an immediate peak, while trusted-node activation may be better for reducing persistence.

This finding complicates conventional crisis communication. Timely and accurate organizational response remains important [6], but corporate response alone may be structurally limited when the firm is not trusted by the audiences most affected by the false claim. A brand may have the most accurate information and still be the wrong messenger. In this sense, brand disinformation defense requires a shift from message-centered crisis response to network-centered credibility management. Reputation is a strategic asset [44], but in a networked environment, that asset is partly held and transmitted by others.

RQ4 asked how algorithmic amplification alters the timing and targeting requirements for containment. The simulations suggest that amplification does not simply increase reach uniformly. It changes which network structures are most vulnerable. In scale-free and influencer-heavy networks, the claim was already near saturation under low amplification, so high amplification had less additional effect. In clustered and polarized networks, amplification had a stronger effect because it helped the claim escape structures that otherwise limited global spread. For brands, this means that a localized narrative can become market-facing not only because users share it, but because platform systems make it more visible.

The paper's contribution follows from this combined view of agents, trust, and networked behavior. System-level outcomes emerge from heterogeneous agent rules, including susceptibility, credibility sensitivity, sharing thresholds, and trust updating [28, 29]. The model incorporates trust directly into the transition process rather than treating it as a post hoc outcome. The main implication is that firms should not think of disinformation preparedness as the ability to issue a fast correction. They should think of it as the ability to prepare audiences, understand network structure, identify credible validators, and intervene before amplification turns a local claim into a broader reputational event.

## VIII. Limitations and Future Work

This paper has several limitations. First, the simulation relies on synthetic assumptions about agent behavior, trust, susceptibility, correction responsiveness, and sharing. This is appropriate for an early computational model intended to examine possible mechanisms rather than predict any one real-world brand attack. Nonetheless, the parameter values are stylized and should be calibrated against empirical data in future work. Agent-based models are most useful when their assumptions are explicit and reproducible [45, 46]. The model should therefore be understood as an initial computational laboratory rather than a final predictive system.

A second limitation is that the model treats the false brand claim as relatively stable. In real online environments, damaging narratives often mutate. A claim may begin as a product allegation, then shift into broader claims about corporate ethics, political values, executive conduct, labor practices, or consumer safety. Future work should extend the model to include narrative mutation, variant dominance, correction failure across variants, and reinfection through modified claims. This would connect the model more directly to open problems in misinformation research, where content, user behavior, propagation, and credibility must often be studied together [4, 5].

A third limitation is that algorithmic amplification is represented through a simplified visibility parameter rather than a full platform recommender system. This abstraction helps isolate increased reach, but does not capture ranking, personalization, feedback loops, or platform-specific exposure dynamics. Future research should incorporate richer models of platform mediation, drawing on work that audits recommender systems and algorithmic exposure in large-scale online platforms [47, 42].

Future work should proceed in three directions. First, the model should be calibrated using observed brand controversies, review fraud, product misinformation, or coordinated reputational attacks in platformed markets [48, 7]. Second, the simulation should be extended to include narrative mutation and variant control. Third, future models should include governance constraints on corporate counter-disinformation, including transparency, disclosure, proportionality, and avoidance of deceptive proxy advocacy. As Epstein notes, one reason to model is to make assumptions visible and testable [27]. The next stage of this work should therefore use empirical data and richer platform models to refine and constrain the mechanisms proposed here.

## IX. Conclusion

We developed and evaluated a multi-agent simulation model of brand-targeted disinformation as a trust-weighted contagion process. We argued that brand disinformation should not be understood only as content classification or conventional crisis communication. A false claim becomes damaging when it moves through networks, reaches agents with different susceptibilities, is reinforced by trusted sources, and is amplified through social and platform mechanisms. This is consistent with work showing that false information can spread differently than true information online [1], and that misinformation requires computational, behavioral, and institutional perspectives [49].

The simulations provide early evidence for this framing. Network structure shaped the extent and speed of spread: scale-free and influencer-heavy networks were especially vulnerable, while clustered and polarized networks provided some structural containment until amplification helped the claim move across communities. Intervention strategy also mattered. Reactive correction reduced peak belief in some conditions, but did not reliably prevent reinfection or persistence. Trusted-node activation was stronger for recovery and containment than for preventing early peaks. Prebunking alone was comparatively weak, but became more effective as part of a hybrid strategy. Across the full simulation set, the hybrid strategy combining prebunking, trusted-node activation, and rapid correction produced the strongest overall results.

These findings reinforce the value of modeling misinformation as a contagion-like process, while also showing why the brand context requires special attention [11]. For firms, reputational harm is produced through interactions among trust, timing, network position, and visibility. Trust is therefore not only an outcome to be repaired after a crisis, it is part of the mechanism by which disinformation spreads or is contained [12]. Similarly, intervention is not merely the issuance of a correction, but the strategic placement of credible counter-information into a live network [34].

The contribution of this paper is a reusable simulation framework for studying brand disinformation defense across alternative network structures, amplification conditions, and response strategies. In doing so, it connects heterogeneous agent behavior, trust-weighted diffusion, and networked information dynamics [29]. In an increasingly connected Web, brand reputation is not held by the firm alone. It is distributed across agents, platforms, communities, and trust relationships. Understanding and defending it requires models that reflect that reality.

## Acknowledgment

Generative AI tools were used to support outlining, organization, drafting, and code development. The author directed research design, source selection, analysis, interpretation, revision, and final manuscript, and retains responsibility for the paper's content.